\documentclass[12pt]{article}

\usepackage{amsmath,amsfonts}
\usepackage{amsthm}
\usepackage[mathscr]{eucal}
\usepackage{epsfig}

\newcommand{\bra}[1]{\pmb{\langle}#1\pmb{|}}

\newcommand{\cdt}{,\ldots,}
\newcommand{\dff}{\scshape}
\newcommand{\dmt}{\rho}
\newcommand{\hhh}{\mathcal{H}}
\newcommand{\hhhh}{\mathcal{H}}
\newcommand{\ket}[1]{\pmb{|}#1\pmb{\rangle}}
\newcommand{\lth}{N}
\newcommand{\otl}{\otimes\cdots\otimes}
\newcommand{\rrh}{\mathbf{\rho}}
\newcommand{\state}{\Psi}
\newcommand{\statee}{\psi}
\newcommand{\yyy}{\mathfrak{s}}
\newcommand{\yyyy}{\mathfrak{S}}

\DeclareMathOperator{\tr}{Tr}

\newcommand{\ketbra}[2]{\ket{#1}\bra{#2}}
 
\newcommand{\rrt}{{\breve{\rrh}}}

\title{A note on Borromean correlations in multipartite quantum 
systems}

\author{
Roman R. Zapatrin\\
\small\itshape 
Quantum Information Group, ISI, 
Villa Gualino,\\ 
\small\itshape 
Viale Settimio Severo 65,
10133, Torino, Italy;\\ 
\small\rm
e-mail: zapatrin@isiosf.isi.it
}

\date{}

\begin{document}

\maketitle

\begin{abstract}
If a pure state of a multipartite quantum system is Borromean, 
that is, such that its density matrix becomes product after tracing 
out any its component then the initial state is product itself. This 
shows the essentially classical nature of Borromean correlations 
which can not be achieved by entangled pure states.

\end{abstract}

A random vector $\left(A_1\cdt{}A_n\right)$ of length $\lth$ 
is said to be {\dff Borromean\footnote{The symbol of three rings 
linked in such a way that taking out any one makes the rest 
disjoint was used in the Renaissance as an heraldic device for the 
Borromeas family of Italy.} correlated} if the distribution 
obtained by ``forgetting'' (or, speaking quantum, tracing out) any 
one of its components is a product one.  As an example of such 
vector when all its components $A_i$ are two-valued consider a 
classical register $(A_0\cdt{}A_\lth)$ of $\lth+1$ bits.  Let each 
bit $A_i\; (i=1\cdt\lth$) takes independently its values 0 and 1 
with probabilities $p_i$ and $q_i$, respectively, while the 0-th 
bit contains the checksum of all the bits: 

\begin{equation}\label{eexborc}
\left\lbrace
\begin{array}{l}
A_1\cdt{}A_\lth\quad\mbox{are independent}\cr
A_0 = A_1\oplus\cdots\oplus{}A_\lth
\end{array}
\right.
\end{equation} 

\noindent Then 
\begin{itemize}
\item the set $(A_0\cdt{}A_\lth)$ of random 
variables is not independent 
\item any marginal distribution 
$(A_0\cdt\widehat{A_i}\cdt{}A_\lth)$ which appears when any bit 
$i=0\cdt\lth$ of the register is traced out is a distribution of 
$\lth$ independent variables.  
\end{itemize}

In classical probability, for a random vector to be Borromean 
correlated, it should necessarily be in a mixed state. It is known 
that in quantum mechanics pure states still can give rise to random 
correlations (which are seen from outside as sharing a secret 
variable \cite{classentang}), and one could expect that there 
exists such a pure quantum state, for which we can write an analog 
of Borromean correlations.  It is shown in this note that no pure 
state of a multipartite quantum system can possess Borromean 
correlated density matrix. 

\medskip 

I consider a composite quantum system $\yyyy$ consisting of $\lth$ 
components $\yyy_r$ each described by its state space 
$\hhh_r,\quad{}r=1,\ldots,\lth$ each of which has dimension $d$. 
The overall state space $\hhhh$ of $\yyyy$ is the tensor product 

\[ 
\hhhh=
\hhh_1\otl\hhh_\lth
\] 

Let $\rrh$ be a density matrix of a state of of $\yyyy$. I call it 
{\dff Borromean} if for any $r=1\cdt\lth$ its partially traced 
(with respect to the subsystem $\yyy_r$) density matrix is product: 

\begin{equation}\label{edefborq}
\rrt=
\tr_r\rrh
=
\dmt_1\otl\dmt_{r-1}\otimes\dmt_{r+1}\otl\dmt_{\lth}
\end{equation}

\noindent The message of this note is contained in the following 

\paragraph{Statement.}
{\em 
Let $\ket{\state}$ be a pure state of the composite 
system $\yyyy$ such that its density matrix 
$\rrh=\ketbra{\state}{\state}$ is Borromean \eqref{edefborq}. Then 
the state $\ket{\state}$ is product.  
}

\medskip 

{\itshape Proof} is essentially based on the generalised Scmidt 
decomposition introduced in \cite{multschmidt}.  Given a state 
$\ket{\state}$, one can always choose such a product basis 
$\{\ket{\statee_i^{(r)}}\mid\;r=1\cdt\lth;\;i=1\cdt{}d\}$ in 
$\hhhh=\hhh_1\otl\hhh_\lth$ that in the decomposition 

\begin{equation}\label{edecpsi}
\ket{\state}=
\sum_{i_1\cdots{}i_\lth}
c_{i_1\cdots{}i_\lth}
\ket{\statee_{i_1}^{(1)}}\cdots
\ket{\statee_{i_\lth}^{(\lth)}}
\end{equation} 

\noindent the coefficients $c_{i_1\cdots{}i_\lth}$ have the 
properties:

\begin{eqnarray}
c_{jii\cdots{}i}=
c_{iji\cdots{}i}=
\cdots=
c_{ii\cdots{}ij}=
0
\quad\mbox{if}\quad
1\le{}i<j\le{}d\label{eprop1}
\\
\left|c_{ii\cdots{}i}\right|\ge
\left|c_{j_1\cdots{}j_n}\right|
\quad\mbox{if}\quad
i\le{}j_r\,;\;r=1\cdt\lth\label{eprop2}
\end{eqnarray}

\noindent Write down the density matrix 

\[
\rrh = 
\ketbra{\state}{\state} = 
\sum_{i_1\cdots{}i_\lth}
\sum_{j_1\cdots{}j_\lth}
c_{i_1\cdots{}i_\lth}
\overline{c}_{i_1\cdots{}i_\lth}
\ket{\statee_{i_1}^{(1)}}\cdots
\ket{\statee_{i_\lth}^{(\lth)}}
\bra{\statee_{i_1}^{(1)}}\cdots
\bra{\statee_{i_\lth}^{(\lth)}}
\]

\noindent and take its partial trace $\rrt=\tr_r\rrh$ with respect 
to the subsystem $\yyy_r$. Then the matrix coefficients 
$\rrt_{i_1\cdots{}i_{r-1}i_{r+1}\cdots{}i_N,\,
j_1\cdots{}j_{r-1}j_{r+1}\cdots{}j_N}$ read: 

\begin{equation}\label{errt}
\rrt_{i_1\cdots{}i_N,\,j_1\cdots{}j_N}=
\sum_{i_r}
c_{i_1\cdots{}i_{r-1}i_r{}i_{r+1}\cdots{}i_\lth}
\overline{c}_{i_1\cdots{}j_{r-1}i_r{}j_{r+1}\cdots{}i_\lth}
\end{equation} 

\noindent Since $\rrh$ is assumed to be Borromean, the density 
matrix $\rrt$ is product, that is, there is a set 
$\rrt^{(1)}\cdt\rrt^{(r-1)},\rrt^{(r+1)}\cdt\rrt^{(N)}$ of 
$d\times{}d$ density matrices in each $\hhh_i$ that 

\begin{equation}\label{err1}
\rrt_{i_1\cdots{}i_N,\,j_1\cdots{}j_N}=
\rrt^{(1)}_{i_1j_1}
\cdot\ldots\cdot
\rrt^{(N)}_{i_Nj_N}
\end{equation} 

\medskip 

If we set all $i_1\cdt{}i_N,\,j_1\cdt{}j_N=1$ then it follows from 
\eqref{eprop1} that all the summands in \eqref{errt} but one 
vanish, namely 
$\rrt_{11\cdots{}11,\,11\cdots{}11}=
c_{11\cdots{}11}\cdot\overline{c}_{11\cdots{}11}>0$, therefore from 
\eqref{err1} we conclude that 

\begin{equation}\label{eallg}
\rrt^{(1)}_{11}>0
\cdt
\rrt^{(N)}_{11}>0
\end{equation} 

\noindent For any $s\neq{}r$, $i_s>1$ it follows from 
\eqref{eprop1} that all summands in \eqref{errt} are 0, therefore we 
infer from \eqref{err1} and \eqref{eallg} that 

\begin{equation}\label{endz}
\forall s\neq{}r,\;
\forall j_s>1
\quad
\rrt^{(s)}_{i_s1}=
\rrt^{(s)}_{1i_s}=
0
\end{equation} 

\noindent Furthermore, for any multi-index $i_1\cdots{}i_N$ ($i_r$ 
excluded) we have 

\[
\rrt_{i_1\cdots{}i_N,\,11\cdots11}=
\rrt^{(1)}_{i_11}
\cdot\ldots\cdot
\rrt^{(N)}_{i_N1}
=0
\]

\noindent if at least one $i_s$ differs from 0. Now consider the 
right-hand side of the expansion \eqref{errt} for this case. Again 
it follows from \eqref{eprop1} that there is at most one non-zero 
summand in \eqref{errt}, that is: 

\[
\rrt_{i_1\cdots{}i_N,\,11\cdots11}=
c_{i_1\cdots{}i_N}
\cdot
\overline{c}_{11\cdots{}11}
\] 

\noindent Since we can repeat our reasoning for any $r$, we have 
proved that for any multi-index $i_1\cdots{}i_N$ ($r$ now included) 
the coefficient $c_{i_1\cdots{}i_N}$ in \eqref{edecpsi} vanishes 
whenever at least one $i_r=1$ and at least one $i_s>1$. 

\medskip 

If $c_{22\cdots22}=0$ then it follows from \eqref{eprop2} that all 
the coefficients in \eqref{edecpsi} are 0 and we are done: 
$\ket{\state}=\ket{\statee_{1}^{(1)}}\cdots
\ket{\statee_{1}^{(\lth)}}$ is a product state. So, suppose 
$c_{22\cdots22}\neq{}0$. In this case, applying \eqref{eprop1} for 
$i=2$ and repeating all the above reasoning we get that for any 
multi-index $i_1\cdots{}i_N$ the coefficient $c_{i_1\cdots{}i_N}$ 
in \eqref{edecpsi} vanishes whenever at least one $i_r=2$ and at 
least one $i_s>2$. Repeating the procedure up to $d$ (the 
dimension of each $\hhh_s$) we see that the only nonzero 
coefficients in the expansion \eqref{edecpsi} are of the form 
$c_{ii\cdots{}ii},\;i=1\cdt{}d$. Now return to the formula 
\eqref{errt} for the reduced density matrix. We see that its only 
non-vanishing elements are of the form 
$\rrt_{ii\cdots{}ii,ii\cdots{}ii}$ which means that it is diagonal. 
Since $\rrt$ is a pure state, $c_{11\cdots11}=1$ and all others 
$c_{ii\cdots{}ii}=0$, therefore 
$\ket{\state}=\ket{\statee_{1}^{(1)}}\cdots 
\ket{\statee_{1}^{(\lth)}}$ is always a product state. 

\medskip 

So, tracing out our pure state $\rrh$ we get a pure {\em product} 
state.  This, in turn, can happen only when the pure state $\rrh$ 
is product itself. 

\paragraph{Concluding remarks.} It was demonstrated that there are 
correlations in quantum  systems which are of purely classical 
nature, that is, they can not be provided  by any pure quantum 
state. 

Note that the condition \eqref{edefborq} for density matrices to be 
Borromean is stronger than its classical analogue. If we weaken the 
condition \eqref{edefborq} and require the Borromean correlations 
not for the density matrix but only for a given fixed set of local 
observables, then it can be achieved by a pure quantum state. This 
can be shown by an example. Consider the following pure state of 
$\lth+1$ qubits: 

\[
\ket{\state}=
\frac{1}{\sqrt{2^\lth}}
\sum_{i_1\cdt{}i_\lth=0,1}
\ket{i_1\oplus\cdt\oplus{}i_\lth}\ket{i_1}\cdots\ket{i_\lth}
\] 

\noindent then the expectation values of the collection of 
observables $\ketbra{1}{1}$ in each qubit has exactly the same 
distribution as shown the example \eqref{eexborc}.

\medskip 

\paragraph{Acknowledgements.} I am grateful to the participants of 
the joint IAKS-ISI workshop (October 4--5, 2001) in particular, to 
Markus Grassl, Dominik Janzing, J\"orn M\"uller-Quade and Martin 
R\"otteler for valuable comments and discussions. The work was 
carried out under the auspices of the EC project Q-ACTA. A support 
from the research grant ``Universities of Russia" is appreciated.

\end{document}